\documentclass[runningheads]{llncs}

\usepackage{amssymb}
\usepackage{subfigure}
\usepackage{graphicx}
\usepackage{graphics}
\usepackage{colortbl}
\usepackage{times}
\usepackage{helvet}
\usepackage{courier}
\usepackage{verbatim}
\usepackage{url}
\usepackage{epstopdf}
\usepackage{algorithm}
\usepackage{algorithmic}
\usepackage{float}
\usepackage{makecell}
\usepackage{multirow}
\usepackage{textcomp}
\usepackage{float}
\usepackage{bm}
\usepackage{hyphenat}
\usepackage{colortbl}
\usepackage{caption}
\usepackage{mdwlist}
\usepackage{CJK}
\usepackage{amsmath}
\usepackage{caption}
\usepackage{subfigure}

\urldef{\mails}\path|yzheng66@iit.edu|

\newcommand{\keywords}[1]{\par\addvspace\baselineskip
\noindent\keywordname\enspace\ignorespaces#1}

\begin{document}

\mainmatter  

\title{MOPO-LSI: A User Guide}

\author{Yong Zheng$^1$, Kumar Neelotpal Shukla$^2$, Jasmine Xu$^2$, David (Xuejun) Wang$^2$, Michael O'Leary$^2$}

\institute{$^1$Illinois Institute of Technology, Chicago, IL, USA 60616\\
$^2$Morningstar, Inc., Chicago, IL, USA 60602\\
Corresponding Email: \mails\\
}

\maketitle

\begin{abstract}
MOPO-LSI is an open-source \underline{M}ulti-\underline{O}bjective \underline{P}ortfolio \underline{O}ptimization \underline{L}ibrary for \underline{S}ustainable \underline{I}nvestments. This document provides a user guide for MOPO-LSI version 1.0, including problem setup, workflow and the hyper-parameters in configurations.

\end{abstract}


\section{Introduction}
\noindent
Several studies~\cite{ponsich2012survey,gunjan2022brief} have examined financial portfolio optimization, which involves selecting financial assets in a meticulous manner to align with an investor's goals while considering their tolerance for risk. Asset allocation, a crucial aspect of this process, entails distributing an investor's portfolio across various asset classes (such as stocks, bonds, and cash) based on their risk tolerance and investment objectives. The main objective of traditional portfolio optimization is to construct a portfolio that either maximizes expected returns at a specific risk level or minimizes risk for a given level of expected returns.

Multiple studies~\cite{renneboog2008socially,busch2016sustainable,talan2019doing} have highlighted the increasing interest in sustainable investments. Investors are recognizing the importance of incorporating Environmental, Social, and Governance (ESG) factors~\cite{friede2015esg,amel2018and} into their decision-making processes. Sustainable investments, also known as socially responsible investing~\cite{renneboog2008socially} or ESG investing~\cite{halbritter2015wages}, involve an investment approach that goes beyond financial returns and considers the impact of investments on the environment, society, and corporate governance practices. This trend signifies a growing awareness among investors about the broader implications of their investment choices.

MOPO-LSI~\cite{zheng2023mopo,compass2023demo} is an open-source \underline{M}ulti-\underline{O}bjective \underline{P}ortfolio \underline{O}ptimization \underline{L}ibrary for \underline{S}ustainable \underline{I}nvestments. It was specifically designed and developed for sustainable investments in mutual funds by using multi-objective optimization (MOO). By integrating ESG factors into portfolio optimization in MOPO-LSI, we aim to generate long-term financial returns while promoting sustainable and responsible practices. This document provides a user guide for MOPO-LSI version 1.0, including problem setup, hyper-parameters and example of the outputs.

\section{MOPO-LSI: Problem Setup}
\noindent
The MOPO-LSI library facilitates asset allocations for a collection of mutual funds, where the dataset provides ESG scores for each mutual fund. These ESG scores are categorized as positive and negative factors. Positive ESG factors encompass aspects beneficial to society, such as clean energy, well-being, and public health. On the other hand, negative ESG variables are associated with harm, such as environmental pollution, regulatory violations, and human rights abuses. In MOPO-LSI, the primary objectives involve constructing a financial portfolio that maximizes positive ESG scores, minimizes negative ESG scores, and manages tracking error. The tracking error quantifies the deviation from standard benchmarks at a specific risk level (e.g., aggressive, moderate, conservative). The risk level serves as a comprehensive factor that captures risks, returns, and volatility. Maintaining the tracking error within a specified range ensures that the portfolio remains within the desired risk level chosen by the user (e.g., aggressive, moderate, conservative).

MOPO-LSI offers portfolio solutions in two scenarios: when client preferences are known and when client preferences are unknown. In the first scenario, client preferences are represented by weights assigned to different objectives. When these preferences are known, the optimization task can be transformed into a single-objective optimization by employing the weighted sum method, as demonstrated in Equation~\ref{eq:weightedsum}.

\begin{equation}
\max (POS_s - NEG_s - p_m\times TE)
\label{eq:weightedsum}
\end{equation}

In the equation above, $s$ refers to a portfolio solution, and $TE$ denotes tracking error (as shown by Equation~\ref{eq:TE}), where $p_m$ refers to the client preference or weight on the tracking error. $POS_s$ and $NEG_s$ represent the weighted sum of ESG scores on positive and negative ESG dimensions, respective, where the weights here refer to client preferences on ESG factors.

 \begin{equation}
POS_s = \frac{\sum\limits_{i\in ESG+}p_i\times Score_i}{\sum\limits_{i\in ESG+}p_i}
\label{eq:poss}
\end{equation}

\begin{equation}
NEG_s = \frac{\sum\limits_{j\in ESG-}p_j\times Score_j}{\sum\limits_{j\in ESG-}p_j}
\label{eq:negs}
\end{equation}

In the equations above, $i$ and $j$ are used to denote a positive and negative ESG factor, respectively. $p_i$ and $p_j$ refer to client preferences on dimension $i$ and $j$. The score shown in Equation~\ref{eq:esg} describes the calculation of ESG score on a single ESG factor or dimension, where $w$ is the vector of fund weights or allocations -- that's the parameter or solution we would like to learn. $E$ denotes the ESG matrix, where each row is a mutual fund, and each column denotes an ESG factor. $E^i$ is used to represent the $i^{th}$ column in $E$.

\begin{equation}
Score_i=\frac{w \cdot E^i}{|w|}
\label{eq:esg}
\end{equation}

\begin{equation}
TE = (w-b)^TV(w-b)
\label{eq:TE}
\end{equation}

In addition to the setup of the objectives, we also assign multiple constraints in order to obtain more feasible and practical solutions:
\begin{itemize}
	\item The summation of fund allocations should be close to 1.
	\item Tracking error must be smaller than a pre-defined threshold which can be easily configured in the library.
	\item Positive ESG scores in the solution should be no smaller than the ones in the benchmark, and negative ESG scores in the solution should be no larger than the ones in the benchmark. The benchmark is determined by a selected risk level.
	\item The portfolio should be diverse, e.g., we cannot assign most of the assets to a same type of the mutual funds. To do so, we can set a threshold for each asset class, and this threshold can be easily defined in the configuration file of the library.
\end{itemize}

By combining the objectives and constraints above, we are able to utilize the embedded conic solver (ECOS)~\cite{domahidi2013ecos} in the CVXPY~\cite{diamond2016cvxpy} library to solve the quadratic convex optimization problem in our library. The output in this scenario will be a single optimal solution.

Alternatively, in cases where client preferences on objectives are not known in advance, we employ various multi-objective evolutionary algorithms (MOEAs)\cite{coello2006evolutionary} from the Pymoo\cite{blank2020pymoo} library to optimize specific goals, which include maximizing positive ESG scores, minimizing negative ESG scores, and tracking error. However, when the number of ESG dimensions increases, MOEAs might require more time to find optimal solutions. To assist clients in identifying their desired solutions more effectively, we allow them to specify individual positive and/or negative ESG factors that they wish to maximize or minimize. This approach enables us to impose constraints on the selected ESG dimensions and ensure that the solutions achieve better ESG scores in those dimensions compared to the benchmark. MOEAs generate a Pareto set, which comprises non-dominated solutions. With MOPO, users can provide additional inputs to employ multi-criteria decision-making methods~\cite{greco2016multiple} and select a single optimal solution from the Pareto set. Furthermore, we offer 2D/3D visualizations to facilitate user observation and understanding of the chosen solution.

\section{MOPO-LSI: Workflow}
\noindent
The workflow by MOPO-LSI can be depicted by Figure 1, with descriptions as follows.
\begin{figure*}[ht!]
\includegraphics[scale=0.65]{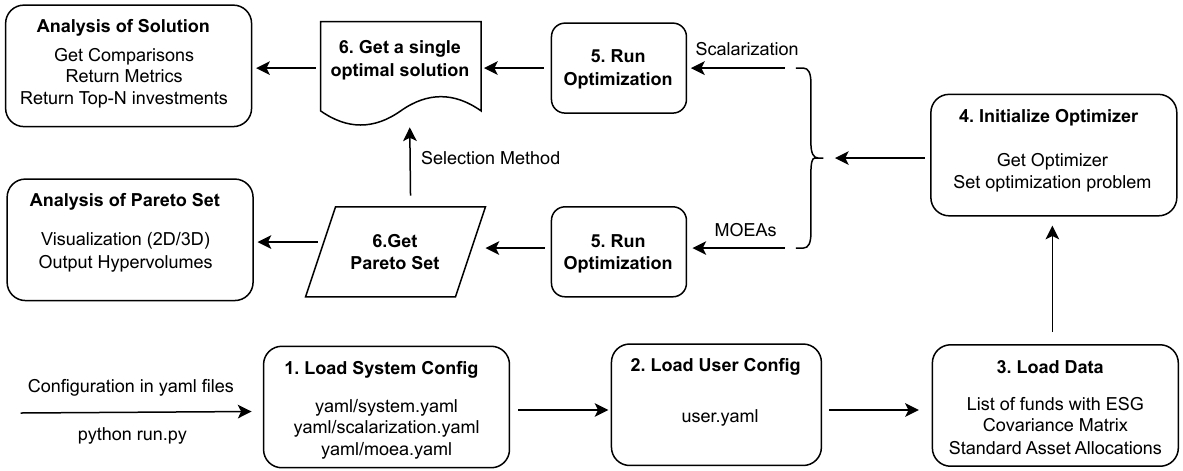}
\caption{The Framework and Workflow in MOPO-LSI~\cite{zheng2023mopo}}
\end{figure*}

\begin{itemize}
	\item \textit{Configurations}. At the beginning, users should well-define hyperparameters in different configuration files, e.g., yaml files. Regarding these parameters, refer to the user guide on the coding repository on Github.
	\item \textit{Run MOPO-LSI}. A user can start running the library by using "python run.py" or "python run.py --config your.yaml". The default yaml to start running is the user.yaml file.
	\item \textit{Load system configurations}. The library will first load system configurations defined by yaml files in the folder "yaml", including system.yaml, scalarization.yaml, moea.yaml. The file "system.yaml" defines file paths, output path, risk levels (i.e., conservative, moderate, aggressive), ESG groups and dimensions, the optimization model to be used, and the general parameters for system constraints. The files, "scalarization.yaml" and "moea.yaml", define the MOO problems and corresponding hyperparameters related to the optimizers.
	\item \textit{Load user configurations}. The file, "user.yaml", is loaded, where the library will read user inputs, such as user option on the risk level, user preferences, and so on.
	\item \textit{Load data}. The library will then load the data sets indicated in "system.yaml", including the list of funds with their ESG scores, the covariance matrix and standard asset allocations for each risk level.
	\item \textit{Initialize optimizer}. The library initializes the optimizer by using the parameter "model" defined in "system.yaml". The MOO problem defined in corresponding yaml, either "scalarization.yaml" or "moea.yaml", will also be set up in this stage. A MOO problem defines the list of objectives and constraints to be considered in the optimization process.
	\item \textit{Run ptimization}. The library will run optimization by using corresponding MOO algorithms (e.g., NSGA2) or optimization solvers (e.g., ECOS for the weighted sum method).
	\item \textit{Get optimization results}. The scalarization methods can return a single optimal solution directly, while the MOEAs will return a Pareto set which is a set of non-dominated solutions. There are several selection methods built in the library to help select a single optimal solution from the Pareto set.
	\item \textit{Analysis of the solutions}. By given a single optimal solution, the library can compare it with the standard benchmark (i.e., a non-ESG optimized solution associated with a specific risk level), output evaluation metrics (i.e., tracking error, improvement ratio on each ESG factor), and finally return the top-$N$ mutual funds for investments. An example of the outputs by using the sample data and the weighted sum method in MOPO-LSI can be observed from Figure 2, where the gain values refer to the weighted average of the improvement ratios over each ESG groups (i.e., positive and negative ESGs). In terms of the analysis for Pareto set, the library has the option to visualize the non-dominated solutions and hypervolumes.
\end{itemize}

\section{User's Guide}
\noindent
In this section, the specific instructions about how to use, deploy and evaluate the MOPO-LSI library.

\subsection{Data Format}
\noindent
A sample data set was provided in the "data/SampleData" folder. Any other practice should follow the format of the following three data files:

\begin{itemize}
	\item funds.csv refers to a list of mutual funds as candidates in portfolio optimization. Each row is a mutual fund with columns as ESG dimensions.
	\item cov\_matrix.csv refers to the covariance matrix of the mutual funds.
	\item asset\_allocation.csv describes benchmark distributions over six asset classes.
\end{itemize}

\subsection{Experimental Configuration}
\noindent
There are several yaml files which store the configurations. 
\begin{itemize}
\item System.yaml defines file paths, output path, risk levels (i.e., conservative, moderate, aggressive), ESG groups and dimensions, the optimization model to be used, as well as the general parameters for system constraints. 
\item The files, "scalarization.yaml" and "moea.yaml", define the MOO problems and corresponding hyperparameters related to the optimizers.
\item The file, "user.yaml", defines user inputs, such as user option on the risk level, user preferences, and so on.
\end{itemize}

More details about these configurations can be found below.

\subsubsection{User.yaml}
\begin{itemize}
\item client\_option: user should declare a specific risk level, i.e., conservative, moderate, or aggressive.
\item client\_preferences: user preferences (for scalarization only) organized for each ESG group along with the strength information (e.g., high, moderate or low). The ESG groups or dimensions with a "high" strength will be assigned the portfolio-level ESG constraints.
\item preloaded\_moea\_results: this option is used for MOEAs only. The non-dominated solutions can be serialized and stored in an external *.pkl file. This pkl file can be loaded into this parameter, so that the library can load these non-dominated solutions directly, and re-select a single optimal solution according to the updated user preferences defined in client\_preferences\_moea.
\item client\_pos\_esg \& client\_neg\_esg: users can declare specific ESG dimensions that are of importance to them. These dimensions will be assigned the portfolio-level ESG constraints.
\item client\_preferences\_moea: simulated client preferences to help select a single optimal solution from Pareto set. 
\end{itemize}

\subsubsection{yaml/Scalarization.yaml}
\begin{itemize}
\item scalar\_problem\_name: a pre-defined MOO problem
\item objectives\_pos\_neg: [True, True], turn on or off optimization for PosESG, NegESG
\item ecos\_eps: tolerance threshold in the ECOS solver
\item ecos\_max\_iter: maximal number of learning iterations in the ECOS solver
\item ecos\_verbose: turn on or off the intermediate outputs from the optimizer
\end{itemize}

\subsubsection{yaml/MOEA.yaml}
\begin{itemize}
\item moea\_problem\_name: a pre-defined MOO problem
\item n\_threads: number of threads used for multi-threading processing
\item n\_population: population size
\item n\_offsprings: number of offsprings
\item n\_generations: number of generations
\item moea\_verbose: turn on or off the intermediate outputs from the optimizer
\item optimal\_selection: ASF or PW
\item output\_visualizations: turn on or off visualization of the single optimal solution
\item output\_visualize\_dims: select two dimensions for visualization, e.g., ['pos\_esg', 'neg\_esg'], you can also add specific ESG dimensions.
\item output\_hypervolume\_visualizations: turn on or off the visualization of hypervolumes. The system runs significantly slowly if it is turned on.

\end{itemize}

\subsubsection{yaml/System.yaml}
\leavevmode\\\\
\noindent
Inputs and Outputs:
\begin{itemize}
\item seed: seed to be used in random process
\item model: an algorithm or solution to run, options: WeightedSum, NSGA2, SMSEMOA, AGEMOEA2
\item dataset: assign the data set name which also refers to the subfolder name in folder "data"
\item data\_funds: funds.csv, the list of funds and ESG scores
\item data\_cov\_matrix: cov\_matrix.csv, the covariance matrix
\item data\_bmk: asset\_allocation.csv, the standard asset allocations 
\item path\_outputs: the output folder where the logs, solutions, visualizations, pkl files will be stored
\item output\_analysis: turn on or off the output analysis, e.g., the output of gains and top-$N$ investments
\item output\_funds\_top\_n: the value of $N$ for top-$N$ investments in the outputs
\end{itemize}

\noindent
General arguments:
\begin{itemize}
\item options: the risk levels, ['Conservative', 'Moderate', 'Aggressive']
\item pos\_esg\_dims: a list of PosESG dimensions, format: dim:category
\item neg\_esg\_dims: a list of NegESG dimensions, format: dim:category
\end{itemize}

\noindent
Constraint arguments:
\begin{itemize}
\item weight\_init\_up\_lim: the up bound for maximal fund allocations
\item weight\_init\_low\_lim: the low bound for maximal fund allocations
\item weight\_final\_low\_lim: the threshold for the library to drop and adjust fund allocations. For example, it is set as 0.001. In the final solution, any funds with allocations smaller than 0.001 will be updated to zero. Fund allocations for other funds will also be re-normalized to make sure the summation of allocation equals to 1.
\item TE\_cap: the cap for tracking errors, not the squared errors
\item adj\_ratio: TE\_cap can be updated to TE\_cap / adj\_ratio
\item esg\_norm\_up\_lim: the up bound for ESG normalizations; it is TE**2, if it is set $\leq$ 0
\item esg\_norm\_low\_lim: the lower bound for ESG normalization, such as 1e-100
\item dev\_asset\_alloc: deviation cap from benchmark allocations for each asset class
\end{itemize}

\section{Acknowledgement}
\noindent
This work was supported and funded by Morningstar, Inc. at Chicago, Illinois, USA, under grant agreement IIT-Cayuse: Grant No. 22-0300, Project No. A23-0011.


\bibliographystyle{abbrv}
\bibliography{sample-base}

\end{document}